\documentclass[english,prb,twocolumn,superscriptaddress,showpacs,preprintnumbers,amsmath,amssymb]{revtex4}

\usepackage{graphicx}% Include figure files
\usepackage{dcolumn}% Align table columns on decimal point
\usepackage{bm}% bold math
\usepackage{epsfig}
\newcommand{\ignore}[1]{}

\begin{document}

\title{Edge Effects on the Electronic Structures of Chemically Modified Armchair Graphene Nanoribbons}

\author{Hao Ren}
\affiliation{Hefei National Laboratory for Physical Sciences at
Microscale, University of Science and Technology of China, Hefei,
Anhui 230026, P.R. China}

\author{Qunxiang Li}
\thanks{Corresponding author. E-mail: liqun@ustc.edu.cn}
\affiliation{Hefei National Laboratory for Physical Sciences at
Microscale, University of Science and Technology of China, Hefei,
Anhui 230026, P.R. China}

\author{Haibin Su}
\affiliation{Division of Materials Science, Nanyang Technological
University, 50 Nanyang Avenue, 639798, Singapore}

\author{Q. W. Shi}
\affiliation{Hefei National Laboratory for Physical Sciences at
Microscale, University of Science and Technology of China, Hefei,
Anhui 230026, P.R. China}

\author{Jie Chen}
\affiliation{Electrical and Computer Engineering, University of
Alberta, AB T6G 2V4, Canada} \affiliation{National Institute of
Nanotechnology, Canada}

\author{Jinlong Yang}
\thanks{Corresponding author. E-mail: jlyang@ustc.edu.cn}
\affiliation{Hefei National Laboratory for Physical Sciences at
Microscale, University of Science and Technology of China, Hefei,
Anhui 230026, P.R. China}

\date{\today}

\begin{abstract}
In this paper, we apply the first-principle theory to explore how
the electronic structures of armchair graphene nanoribbons (AGNRs)
are affected by chemical modifications. The edge addends include H,
F, N, NH$_{2}$, and NO$_{2}$. Our theoretical results show that the
energy gaps are highly tunable by controlling the widths of AGNRs
and addends. The most interesting finding is that N-passivated AGNRs
with various widths are metallic due to the unique electronic
features of N-N bonds. This property change of AGNRs (from
semiconducting to metallic) is important in developing
graphene-based devices.
\end{abstract}


\maketitle

\section{INTRODUCTION}

Graphene, a single atomic layer of graphite with a honeycomb crystal
structure, has attracted a great number of research
activities.\cite{NatureMater07} Its structural properties,
electrical conductance and quantum Hall effects have been
investigated for making novel nanoelectronic
devices.\cite{Science04-Geim,Nature05-Geim,Nature05-Kim} Due to the
linear energy dispersion relation near the Dirac points, graphene is
an interesting conductor in which electrons move like massless
Dirac-fermions.\cite{Nature05-Geim} It is now possible to make
graphene nanoribbons (GNRs) with various experimental methods such
as tailoring via a scanning tunneling microscopy
tip,\cite{ASS04-Hiura} exfoliating from high oriented pyrolytic
graphite,\cite{Nature05-Geim,Nature05-Kim,Science07-Bunch}, or
graphitizating SiC wafers.\cite{Science06-Berger} The energy gaps of
GNRs with several tens of nanometers can be measured because the
growth of the energy gap is inversely proportional to the GNRs'
width.\cite{PRL-Han} There are two typical types of GNRs according
to their edge configurations, either armchair or zigzag. Zigzag edge
GNRs are metallic without considering the freedom of spin because
the two edge states are degenerated at the Fermi level.
Hydrogen-saturated armchair GNRs (AGNRs), on the other hand, are
semiconductors.\cite{PRL-Nature-Louie,PRB-Zheng} The electronic
transport and magnetic properties of zigzag GNRs have also been
studied by several research
groups.\cite{PRL-Nature-Louie,NL-HM,APL-HM, JCP-Jiang}

Because of the high ratio between edge and inner atoms, the
electronic structure of narrow GNRs are sensitive to various edge
addends. Individual carbon atoms on graphene edges are only bounded
to two neighboring carbon atoms and a dangling carbon bond offers a
remarkable opportunity for altering GNR's electronic properties.
This edge modification can be implemented by attaching various atoms
or functional groups to these dangling carbon atoms. To the best of
our knowledge, edge modified AGNRs, except for the
hydrogen-passivated GNRs, have not been systematically examined
using Density Functional Theory (DFT). In this paper, the electronic
structures of AGNRs, with various addends including H, F, N,
NO$_{2}$ and NH$_{2}$, are thoroughly investigated based on
first-principle calculations. Our results show that AGNRs can be
either semiconducting or metallic by changing edge chemical addends.
This remarkable characteristic is very useful in making
graphene-based molecule computing or sensing devices. The detailed
analysis of the electronic structures can also be used to understand
the underlying microscopic mechanisms of these edge effects.

\section{MODEL AND METHOD}

The electronic structure calculations are carried out by employing
the Vienna \textit{ab initio} simulation package with local density
approximation.\cite{vasp,LDA} The electron-ion interactions are
described based on the projected augmented wave (PAW) and the frozen
core approximation.\cite{PAW} The energy cutoff is set to be 400 eV.
Following the previous definition of AGNRs,\cite{PRL-Nature-Louie}
we choose AGNRs with width W=18, 19, and 20 (corresponding to 3p,
3p+1, and 3p+2, respectively. Here, p is an integer) as examples to
study three kinds of AGNRs. The schematic of a sample AGNR with
width of W=18 is shown in Figure 1. Various addends, such as H, F,
and N atoms, are connected to individual edge carbon atoms. In the
NH$_{2}$ and NO$_{2}$ passivated cases, each edge carbon atom bonds
to one NH$_{2}$ or NO$_{2}$ group alternatively (its neighboring
carbon atom is saturated by an H atom). In our calculations, all
atomic positions are allowed to relax. The convergence tolerance in
energy and force is 2$\times$10$^{-5}$ eV and 0.02 eV/{\AA},
respectively. The vacuum layers between two neighboring AGNR sheets
and AGNR edges are set to be $12$ {\AA} and $15$ {\AA} wide,
respectively. In a typical calculation, a one-dimensional periodic
boundary condition along the edge direction is imposed. We choose
the k-point sampling consisting of $15$ uniform k-points together
with the $\Gamma$ point.

\begin{figure}
\includegraphics[width=8.5cm]{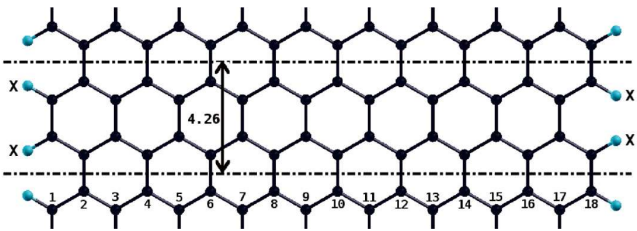}
\caption{(Color online) Schematic of a sample AGNR with width W=18.
The black dots correspond to carbon atoms while the blue ones
represent the addends (labeled with X, where X can be H, F, N, 2H,
2F, NO$_{2}$, and NH$_{2}$ molecules). The distance between two
dash-dotted lines is the lattice constant of one-dimensional
unit-cell in the DFT calculations.}
\end{figure}

\section{RESULTS AND DISCUSSIONS}

We first choose the well-studied hydrogen passivated
AGNRs\cite{PRL-Nature-Louie,PRB07-Wang} as a basis to validate our
numerical results. The calculated band structures of three kinds of
hydrogen passivated AGNRs are plotted in Figure 2 (a). The
corresponding bandgaps together with the optimized C-C distances at
the edge are listed in Table I. Our results show that all three
kinds of AGNRs are semiconductors with direct band gaps at the
$\Gamma$ point. The figure clearly shows that the AGNR with width
$W=3p+1$ has the largest energy gap, while the AGNR with width
W=3p+2 has the smallest one. Our calculated bandgaps are 0.41, 0.61,
and 0.10 eV for the H passivated AGNRs with width W=18, 19 and 20,
respectively. These results agree well with the previously reported
results \cite{PRL-Nature-Louie,PRB07-Wang}.

\begin{table}
\caption{The calculated energy gaps (in eV) of X-AGNRs with width
w=18, 19, and 20. Here, ``M" stands for the metallic case without
E$_{gap}$.} \label{Table 1}
\begin{tabular}{l}
\begin{tabular}{lllllllll}
\hline \hline
W & X=             & H & F & 2H & 2F & NO$_{2}$ & NH$_{2}$ & N \\
\hline
18 & E$_{gap}$(eV) & 0.41 & 0.61 & 0.70 & 0.45 & 0.38 & 0.47 & M \\
19 & E$_{gap}$(eV) & 0.61 & 0.35 & 0.10 & 0.15 & 0.57 & 0.38 & M \\
20 & E$_{gap}$(eV) & 0.10 & 0.14 & 0.42 & 0.63 & 0.09 & 0.13 & M \\
\hline \ &$\Delta E_{f}(eV)\footnote{Because there is not much
difference in bandgaps when narrow graphenes are selected, only the
formation energy of AGNRs with width w=18 is considered. The
formation energy is defined as, $\Delta E_{f}={1\over n}
(E_{tot}-E_{bare}-\sum n_{X}\mu_{X})$, where $E_{tot}$, $E_{bare}$,
n, $n_{X}$, and $\mu_{X}$ is the total energy of system, the energy
of unpassivated 18-AGNR, the number of chemical groups, the number
of atoms of addends X and the chemical potential of X, respectively.
We choose the chemical potential of H, F, N, and O as the binding
cohensive energy per atom in H$_{2}$, F$_{2}$, N$_{2}$, and O$_{2}$
molecules. This definition would be a appropriate mesurement of the
stability in passivated AGNRs. The formation energy of NO$_{2}$ and
NH$_{2}$ passivated 18-AGNRs per group is defined as $\Delta
E_{f}=\frac{1}{2}(E_{tot}-E_{bare}-\sum n_{X}\mu_{X}-2\times\Delta
E_{f}(H))$, where $\Delta E_{f}(H))$ is the formation energy of
H-passivated 18-AGNR.}$
 &-2.29&-3.78&-1.35&-3.06&-4.05&-2.77 &-1.51\\
 \hline \hline
\end{tabular}\\
\end{tabular}
\end{table}

\begin{figure}
\includegraphics[width=8.5cm]{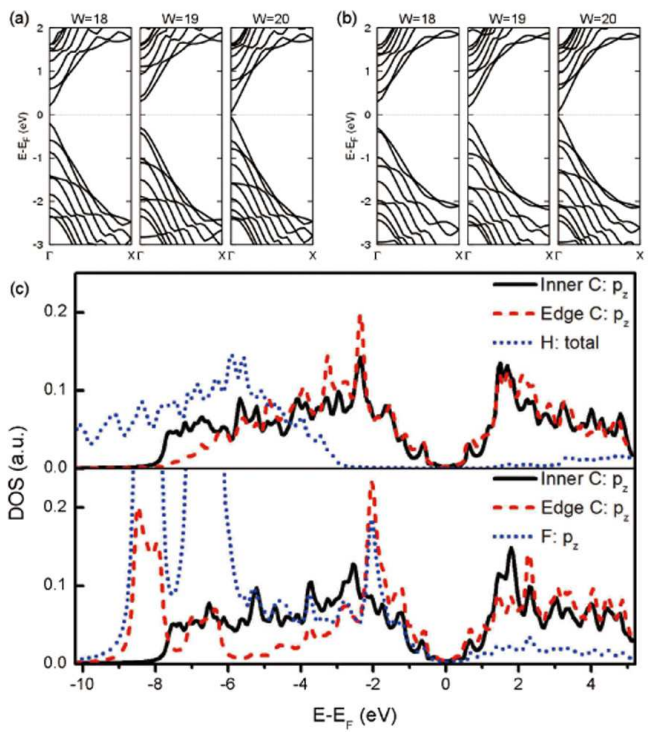}
\caption{(Color online) (a) and (b) show the band structures of H
and F-passivated AGNRs with width W= 18, 19, and 20, respectively.
(c) The projected density of states(PDOS) of H(upper panel) and
F(lower panel) passivated AGNRs with width W=18.}
\end{figure}

Next, we examine the electronic structures of AGNRs with F atoms
passivated at both edges. The calculated band structures are plotted
in Fig. 2 (b) and the bandgaps are listed in Table I. Similar to the
H-passivation cases, all F-passivated AGNRs are semiconductors.
However, the effects of GNR width on bandgaps is clearly different.
For instance, the F-passivation makes the band gap of the AGNR with
width W=18 (3p) to be the largest (0.61 eV), while the gap of the
AGRN with width W=20 (3p+2) is the smallest (0.14 eV). It is
important to note that the bandgaps of the AGNRs (width W=20) with
the smallest gaps remain as the smallest even after passivation.
However, AGNRs with the largest bandgaps behave differently and
depend on elements used for passivation. We also would like to know
what happens if individual edge carbon atoms connect to two H or F
atoms instead of one H or F atom. Intuitively, the sp$^{2}$
hybridization of edge carbon atoms is converted into sp$^{3}$
hybridization. The effective AGNR width is therefore reduced by 2.
In other words, the effective width becomes $17$ for both double H-
and F-passivated AGNRs when their initial width is W=19. Because the
effective width of $17$ fits 3p+2, the corresponding double H- and
F-passivated AGNRs have the smallest energy gap (about 0.1 eV). The
largest gaps are 0.70 eV for double H-passivated AGNRs with W=17 or
effective W=16 (3p+1). Similarly, $0.63$ eV for of double
F-passivated AGNRs with W=20 or effective W'=18 (a 3p case). These
results are consistent with the width dependent energy gap features
in single F- or H-passivated AGNRs.

It is instructive to compare the DFT results with the tight-binding
approximation (TBA) data. Wang et al. have reported that energy gaps
can be tuned by changing edge hopping parameters.\cite{PRB07-Wang}
Son \emph{et al.} have also showed that the energy gaps of
H-passivated AGNRs can be reproduced via TBA simulations. In their
simulations, edge hopping parameters can be increased by decreasing
the bond length of C atoms at edges.\cite{PRL-Nature-Louie} Although
the geometric deformations of carbon networks are almost the same
between H-passivated GNRs and F-passivated ones, the bandgaps
surprisingly show distinct dependence on the width of AGNRs.
According to the TBA parameters adopted in these analytical
calculations of the edge modified GRNs,\cite{Porezag} the $\pi_{pp}$
hopping integrals, representing the coupling between $p_{z}$ orbital
of two neighboring carbon atoms, increases as the C-C distance
decreases. However, this statement is invalid when the bond length
substantially decreases so that the hybridization between two carbon
atoms becomes perturbed. From the PDOS of the H-passivated AGNRs, we
can see that the peaks of $p_{z}$ orbital of edge C atoms almost
coincide with that of the inner C atoms below the Fermi level {[}see
Fig2(c)]. The $\pi_{pp}$ bonding remains unchanged during the
H-passivation of edge C atoms. The modification of energy gaps
caused by H atoms could be explained based on the geometry
deformation. Interestingly, the $p_{z}$ orbital of both edge C atoms
and F addends have peaks at $-8.5$ eV below the Fermi energy. The
peak position of the edge C atoms, however, decreases significantly
compared to the inner C atoms from $-6.0$ to $-2.3$ eV below the
Fermi energy, which is still within the energy range of $\pi$
bonding in F-passivated AGNRs. That is to say, the $p_{z}$ orbital
of edge C atoms in the F-passivated cases form chemical bonds with
the neighboring C atoms and the F atoms. Consequently, the hopping
parameters decrease. If we choose the hopping parameter to be 84\%
of -2.7 eV ( the typical hopping integral of sp$^{2}$ carbon
system), the energy gaps of AGNRs with width W=18, 19, and 20 are
0.65, 0.35, and 0.13 eV, respectively, which agree excellently with
results obtained by DFT. This observation demonstrates that the
electronic structures of AGNRs can be controlled by edge chemical
modifications.

\begin{figure}
\includegraphics[width=8.5cm]{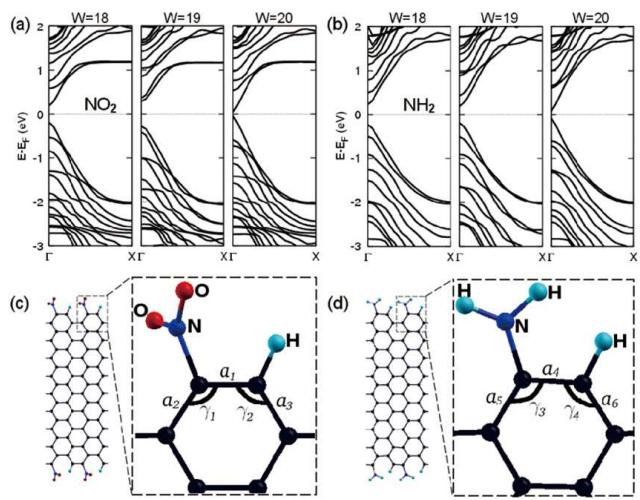}
\caption{(Color online) Band structures of NO$_{2}$-, and
NH$_{2}$-passivated AGNRs with width W = 18, 19, and 20 are show in
(a) and (b), respectively. The edge relaxed geometric structures of
NO$_{2}$ and NH$_{2}$ are shown in (e) and (f). The relaxed bond
lengths from $a_{1}$ to $a_{6}$ are 1.36, 1.40, 1.39, 1.37, 1.42,
and 1.37 {\AA}, respectively, and the bond angles from $\gamma_{1}$
to $\gamma_{4}$ are 122.8$^{\circ}$, 120.6$^{\circ}$,
117.5$^{\circ}$, and 126.6$^{\circ}$, respectively.}
\end{figure}

\begin{figure}
\includegraphics[width=8.5cm]{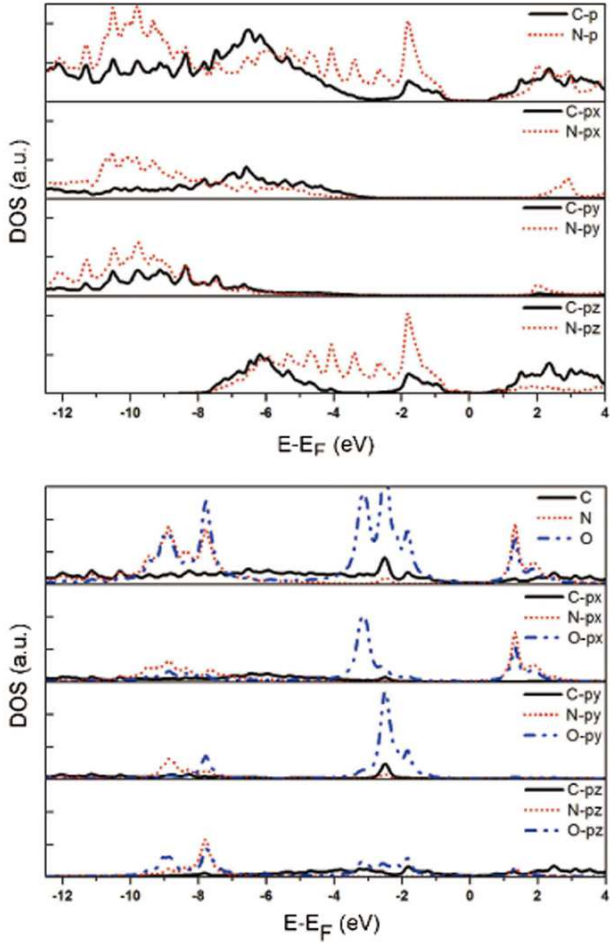}
\caption{(Color online) (a) Partial density of states (PDOS) of the
edge C, and N in NH$_{2}$-passivated AGNRs. (b) PDOS of the C, N,
and O atoms in NO$_{2}$-AGNRs. Here, W=18.}
\end{figure}

NO$_{2}$ and NH$_{2}$ radicals are widely used as anchoring or side
groups for making single molecular junctions in molecular
electronics. \cite{Venkataraman1,Venkataraman2,Star} In the next
calculations, NO$_{2}$ and NH$_{2}$ are attached to AGNRs and we
investigate how these addends impact GNR electronic structures.
Figs. 3 (a) and (b) show their band structures.  Figs. 3 (c) and (d)
show their optimal geometric structures. The NH$_{2}$ group lies in
the AGNR plane for NH$_{2}$ passivated GNRs, while, for NO$_{2}$
passivated GNRs, the dihedral angle is 59$^{\circ}$ between AGNR and
NO$_{2}$ plane. According to the projected DOS in Figure 4, their
electronic structures also exhibit different features. For the
NH$_{2}$-passivated AGNRs, the nitrogen atom forms one sp$^{2}$
orbital, which bonds to two hydrogen and one edge carbon atom. The
remaining p$_{z}$ orbital of the nitrogen atom bonds nicely to the
p$_{z}$ orbital of the edge carbon. For the NO$_{2}$ case, the $\pi$
orbital in the NO$_{2}$ plane does not bond well to the sp$^{2}$
framework of AGNRs because the NO$_{2}$ plane twists out of the AGNR
sheet. Comparing to the bandgaps of H- and F-passivated AGNRs, it is
interesting to note that the bandgap of NO$_{2}$ is close to that of
H, while the bandgap of NH$_{2}$ is similar to that of F. The
fundamental cause is attributed to the influences of the carbon
p$_{z}$ orbital. Neither H nor NO$_{2}$ forms a bond with the carbon
p$_{z}$ orbital, but both F and NH$_{2}$ facilitate bonding directly
to carbon p$_{z}$ orbital. Hence, it is expected that substantial
charge transfer occurs together with the non-negligible change in
the hopping integral near the edges of both F- and
NH$_{2}$-passivated AGNRs.

\begin{figure}
\includegraphics[width=8.5cm]{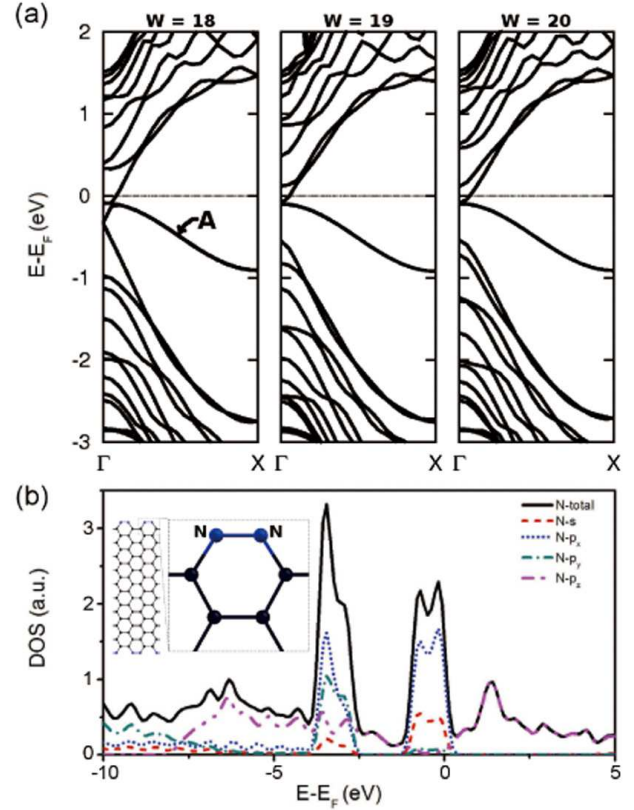}
\caption{(Color online) (a) Band structures and DOS of N-passivated
AGNRs with width W = 18, 19, and 20, respectively. Here, the total
DOS and projected DOS on N atoms are labeled with solid and dashed
lines. (b) Partial projected DOS on N atoms of N-passivated AGNRs
with width W=18. The relaxed geometry is shown in the inset in
Fig.(b)}
\end{figure}

Very recently, N-rich Oligoacences were considered to be candidates
for n-channel organic semiconductors.\cite{Winkler} N-doped carbon
nanotubes and nanofibers have also been synthesized and investigated
intensively over the past years.\cite{Terrones, ZWS, Harigaya,
Yoshioka} These investigations motivate us to examine the
N-passivated AGNRs. According to the calculated formation energy in
Table 1, it is possible to attach one N atom to each C atom along
the edges as an addend. Interestingly, two neighboring N addends can
form an N-N dimer and form an optimal structure. The N-N bond length
is 1.30 {\AA}, which is 0.20 {\AA} longer than that of N-N bond
length in N$_{2}$ molecule. More importantly, a
semiconductor-to-metal transition is observed for all N-passivated
AGNRs and this transition is independent of AGNR widths. To explore
the cause at the microscopic scale, the band structures of
N-passivated AGNRs with various graphene widths and partial
projected DOS on N atoms of AGNR with width W=18 are presented in
Figs. 5(a) and 5(b), respectively. Clearly, the DOS near the Fermi
level is impacted by N addends. The edge N atom has 5 valence
electrons. Two of them stay as a lone-pair, which locates in the
AGNR' plane with the direction pointing outwards. The occupied band
labeled with {}``A\char`\"{} in Fig. 5(a) has the same features as
those of lone-pair electrons. The other three electrons exhibit
sp$^{2}$-like hybridization. Two of the sp$^{2}$ hybridized orbital
pair with the neighboring C and N atoms to form two $\sigma$ bonds,
while the last electron acts as a $\pi$ electron. We would like to
emphasize that the metallic state of N-passivated AGNRs is
predominantly contributed by the edge states of N addends. There are
two reasons attributed to AGNR's semiconducting to metallic
transition. i) the N-N bond is mainly composed of 2s and $2p_{y}$
orbitals. The energy of the band with N $2p_{z}$ is above the Fermi
surface, which means that N $2p_{z}$ electrons fill the carbon
$\pi*$ bands and push the Fermi surface inside the carbon's $\pi*$
bands. ii) the lone-pair electrons of N atoms are mainly in N
$2p_{x}$ bands and the top part of $2p_{x}$ bands pass the Fermi
surface. This feature contributes to the semiconducting to the
metallic transition. Clearly, edge N-N addends extend the sp$^{2}$
network of AGNRs and thus the effective width (or the number of
sp$^{2}$ dimer lines) of N-passivated AGNRs increases by 2 (coming
from two edge sides). Apart from {}``A\char`\"{} band, the band
structures of N-passivated AGNRs with width W are similar to those
of H or F-passivated AGNRs with width $W+2$ except that the Fermi
level shifts upwards and cross the lowest unoccupied bands, which
leads to the semiconductor-to-metal transition.

\section{Conclusion}

In summary, we perform the first-principle calculations on the
electronic structures of AGNRs with various edge chemical
modifications. We find that the energy gap can be tuned to be either
metallic or semiconducting by attaching different chemical
functional addends. After AGNRs are passivated by H, F, NH$_{2}$,
and NO$_{2}$, the GNRs with the smallest bandgaps (i.e. W=20) remain
to have the smallest bandgaps. However, AGNRs that have the largest
bandgaps behave differently and the gaps depend on the type of
elements used for passivation. H and F have similar bandgaps to
NO$_{2}$ and NH$_{2}$, respectively, which can be explained based on
the bonding of the carbon p$_{z}$ orbital. The fascinating
observation is that AGNRs passivated by attaching N atoms can change
from semiconducting to metallic material regardless of AGNR width.
This result is analyzed in detail based on its electronic structure.
The unique electronic structure of the N-N bond, which is
responsible for semiconductor-to-metal transition, is explained
based on its 2s and $2p_{y}$ orbital, and the lone-pair of $2p_{x}$.
This tunable character of AGNRs is useful in developing
graphene-based molecule electronic devices in the near future.

\section*{ACKNOWLEDGMENTS}

This work was partially supported by the National Natural Science
Foundation of China under Grant Nos. 20773112, 10574119, 50121202,
and 20533030, by National Key Basic Research Program under Grant No.
2006CB922004, by the USTC-HP HPC project, and by the SCCAS and
Shanghai Supercomputer Center.Work at NTU is supported in part by a
COE-SUG grant (No. M58070001) and A{*}STAR SERC grant (No.
0521170032). Jie Chen would like to acknowledge the funding support
from the Discovery program of Natural Sciences and Engineering
Research Council of Canada under Grant No. 245680.

\end{document}